\documentclass[12pt]{article}
\usepackage{graphicx}
\usepackage{endfloat}
\begin{document}
\title{ Realization of generalized quantum searching using nuclear magnetic resonance}
\author{\small{} Jingfu Zhang$^{1}$ , Zhiheng Lu$^{1}$,Lu Shan$^{2}$,and Zhiwei
Deng$^{2}$  \\
\small{} $^{1}$Department of Physics,\\
\small{}Beijing Normal University, Beijing,
100875, Peoples' Republic of China\\
\small{} $^{2}$Testing and Analytical Center,\\
\small{}  Beijing Normal University,
 Beijing,100875, Peoples' Republic of China}
\date{}
\maketitle
\begin{minipage}{120mm}
\hspace{0.5cm}
{\small
  According to the theoretical results, the quantum
searching algorithm can be generalized by replacing the
Walsh-Hadamard(W-H) transform by almost any quantum mechanical
operation. We have implemented the generalized algorithm using
nuclear magnetic resonance techniques with a solution of
chloroform molecules. Experimental results show the good agreement
between theory and experiment. }

 PACS number(s):03.67,89.70
\end{minipage}
\vspace{1.5cm}

 The quantum searching algorithm was first
proposed by Grover [1]. It can speed up some search applications
over unsorted data and has been become a hot topic in quantum
information [2$\sim$5]. The surprising result of the algorithm is
that searching a particular item in unsorted list of N elements
only requires $O(\sqrt{N})$ attempts [2]. The algorithm needs the
properties of Walsh-Hadamard(W-H) transform and has been
implemented using NMR techniques [6]. Grover generalized the
algorithm in theory, in which the Walsh-Hadmamard transform can be
replaced by almost any quantum operation [7].

The generalized algorithm can be posed as follows. If a unitary
operator U is applied to a quantum system in an initial basis
state $|\gamma>$, the system lies in a superposition $\sum_{i}
a_{i}|i>$, where $a_{i} = <i|U|\gamma>=U_{i\gamma}$. The amplitude
of reaching the target state $|\tau>$ is $U_{\tau\gamma}$, and the
probability of getting the system in state $|\tau>$ is
$|U_{\tau\gamma}|^{2}$ if a measurement is made. Therefore, it
will take at least $O(1/|U_{\tau\gamma}|^{2})$ repetitions of U
and the measurement to get state $|\tau>$ for one time. In
contrary, according to the generalized searching algorithm, in
$O(1/|U_{\tau\gamma}|)$ applications, operator Q transforms
$|\gamma>$ into $U^{-1}|\tau>$. After U is applied, the system
lies in state $|\tau>$. Q is defined as
$Q\equiv-I_{\gamma}U^{-1}I_{\tau}U$, where $I_{i}\equiv
I-2|i><i|$, and I denotes unit matrix. If $I_{i}$ is applied to a
superposition of states, it only inverts the amplitude in state
$|i>$, and leaves the other states unaltered. It has been showed
in theory that the number of repetitions of Q required to
transform $|\gamma> $ into $|\tau>$ is $\pi/4|U_{\tau\gamma}|$ if
$|U_{\tau\gamma}|\ll1$.

  It can be proved that for a two qubit system, the
number of repetitions of Q is also $O(1/|U_{\tau\gamma}|)$ without
the condition that $|U_{\tau\gamma}|\ll1$. Particularly, if
$|U_{\tau\gamma}|=1/2$, the number of repetitions is 1, and the
probability of getting the target state is 1. In this paper, we
will implement the generalized quantum algorithm in this case on a
two qubit NMR quantum computer [8].

Our experiments use a sample of Carbon-13 labelled chloroform
(Cambridge Isotopes) dissolved in d6-acetone. Data are taken at
controlled temperature (22$^{0}C$) with a Bruker DRX 500 MHz
spectrometer (Beijing Normal University). The resonance
frequencies $\nu_{1}=125.76$ MHz for $^{13}C$, and
$\nu_{2}=500.13$ MHz for $^{1}H$. The coupling constant $J$ is
measured to be 215 Hz. If the magnetic field is along
$\hat{z}$-axis, and let $\hbar=1$, the Hamitonian of this system
is
\begin{equation}\label{1}
  H=-2\pi\nu_{1}I_{z}^{1}-2\pi\nu_{2}I_{z}^{2}+2\pi J I_{z}^{1}
  I_{z}^{2},
\end{equation}
where $I_{z}^{k}(k=1,2)$ are the matrices for $\hat{z}$-component
of the angular momentum of the spins [9]. In the rotating frame of
spin k, the evolution caused by a radio-frequency(rf) pulse along
$\hat{x}$ or $\hat{y}$-axis is denoted as $ X_{k}(\varphi_{k})=
e^{i\varphi_{k}I_{x}^{k}}$ or $ Y_{k}(-\varphi_{k})=
e^{-i\varphi_{k}I_{y}^{k}}$, where
$\varphi_{k}=B_{1}\gamma_{k}t_{p}$. $B_{1} $, $\gamma_{k}$ and
$t_{p}$ represent the strength of magnetic field, gyromagnetic
ratio of spin k and the width of the rf pulse, respectively. The
pulse used above is denoted as $[\varphi]_{x}^{k}$ or
$[-\varphi]_{y}^{k}$.
 The coupled-spin evolution is denoted as
\begin{equation}\label{2}
  [t]=e^{-i2\pi J I_{z}^{1} I_{z}^{2}},
\end{equation}
where t is evolution time. The pseudo-pure initial state
\begin{equation}\label{3}
  |\gamma>=|\uparrow>_{1}|\uparrow>_{2}=\left(\begin{array}{c}
    1 \\
    0 \\
    0 \\
    0 \
  \end{array}\right)
\end{equation}
is prepared by using spatial averaging [10], where
$|\uparrow>_{k}$ denotes the state of spin k. For convenience, the
notion $|\uparrow>_{1}|\uparrow>_{2}$ is merged into
$|\uparrow\uparrow>$ and the subscripts 1 and 2 are omitted. The
basis states are arrayed as
$|\uparrow\uparrow>,|\uparrow\downarrow>,|\downarrow\uparrow>$, $
|\downarrow\downarrow>$. In matrix notion, $I_{i}$ is a diagonal
matrix with all diagonal terms equal to 1 except the $ii$ term
which is -1. For example, if $|i>=|\downarrow\downarrow>$, $I_{i}$
is expressed as
\begin{equation}\label{4}
  I_{3}=\left(\begin{array}{cccc}
    1 & 0 & 0 & 0 \\
    0 & 1 & 0 & 0 \\
    0 & 0 & 1 & 0 \\
    0 & 0 & 0 & -1 \
  \end{array}\right).
 \end{equation}
 U is chosen as three forms
\begin{equation}\label{5}
  U_{1}=1/2\left(\begin{array}{cccc}
    -1 & -1 & -1 & -1 \\
    -1 & 1 & -1 & 1 \\
    -1 & -1 & 1 & 1 \\
    -1 & 1 & 1 & -1 \
  \end{array}\right),
\end{equation}
\begin{equation}\label{6}
  U_{2}=1/2\left(\begin{array}{cccc}
    1 & 1 & 1 & 1 \\
    -1 & 1 & -1 & 1 \\
    -1 & -1 & 1 & 1 \\
    1 & -1 & -1 & 1 \
  \end{array}\right)
\end{equation} and
\begin{equation}\label{7}
  U_{3}=1/2\left(\begin{array}{cccc}
     1 & i & i & -1 \\
     i & 1 & -1 & i \\
     i & -1 & 1 & i \\
    -1 & i &  i &  1 \
  \end{array}\right),
\end{equation}
where $U_{1}$ is the W-H transform used in reference [6] (up to an
phase factor). It is easily proved that $UQ$ can transform state
$|\gamma>$ into state $|\tau>$. The irrelevant overall phase
factors are ignored.

 The evolution of the system can be
represented by its deviation density matrix $\rho_{\Delta}$[11].
For the heteronuclear system, the following rf and gradient pulse
sequence
$[\alpha]_{x}^{2}-[grad]_{z}-[\pi/4]_{x}^{1}-1/4J-[\pi]_{x}^{1}-[\pi]_{x}^{2}-1/4J-[-\pi/4]_{y}^{1}-[grad]_{z}$
transforms the system from the equilibrium state
\begin{equation}\label{8}
  \rho_{\Delta eq}=\gamma_{1} I_{z}^{1}+\gamma_{2} I_{z}^{2}
\end{equation}
to the initial state required in experiments
\begin{equation}\label{9}
  \rho_{\Delta 0}=-(I_{z}^{1}/2+I_{z}^{2}/2+I_{z}^{1}I_{z}^{2})=-1/4\left(\begin{array}{cccc}
    3 & 0 & 0 & 0 \\
    0 & -1 & 0 & 0 \\
    0 & 0 & -1 & 0 \\
    0 & 0 & 0 & -1 \
  \end{array}\right),
\end{equation}
where $\alpha=\arccos(\gamma_{1}/2\gamma_{2})$, and $[grad]_{z}$
denotes gradient pulse along $\hat{z}$-axis. The pulses are
applied from left to right. The symbol 1/4J denotes the evolution
caused by the Hamitonian H for 1/4J without pulses. The initial
state can be used as the pseudo-pure state $|\uparrow\uparrow>$ in
NMR quantum computation [12]. By applying pulse $[\pi]_{x}^{2}$,
$[\pi]_{x}^{1}$
 or $[\pi]_{x}^{1}[\pi]_{x}^{2}$, the pseudo- pure states
corresponding to states $|\uparrow\downarrow>$,
$|\downarrow\uparrow>$ or $|\downarrow\downarrow>$ can be gotten.
They are written as
\begin{equation}\label{10}
 \rho_{\Delta 1}=-(I_{z}^{1}/2-I_{z}^{2}/2-I_{z}^{1}I_{z}^{2})
\end{equation}
\begin{equation}\label{11}
 \rho_{\Delta 2}=-(-I_{z}^{1}/2+I_{z}^{2}/2-I_{z}^{1}I_{z}^{2})
\end{equation}
\begin{equation}\label{12}
 \rho_{\Delta 3}=-(-I_{z}^{1}/2-I_{z}^{2}/2+I_{z}^{1}I_{z}^{2})
\end{equation}
We also represent the four pseudo- pure states above as
$|\uparrow\uparrow>,|\uparrow\downarrow>,|\downarrow\uparrow>,$
and $|\downarrow\downarrow>$, respectively. According to
reference[6], we find that

$I_{0}=Y_{1}(\pi/2)Y_{2}(\pi/2)X_{1}(-\pi/2)X_{2}(-\pi/2)Y_{1}(-\pi/2)Y_{2}(-\pi/2)[1/2J]$,

$I_{1}=Y_{1}(\pi/2)Y_{2}(\pi/2)X_{1}(\pi/2)X_{2}(-\pi/2)Y_{1}(-\pi/2)Y_{2}(-\pi/2)[1/2J]$,

$I_{2}=Y_{1}(\pi/2)Y_{2}(\pi/2)X_{1}(-\pi/2)X_{2}(\pi/2)Y_{1}(-\pi/2)Y_{2}(-\pi/2)[1/2J]$
 and

$I_{3}=Y_{1}(\pi/2)Y_{2}(\pi/2)X_{1}(\pi/2)X_{2}(\pi/2)Y_{1}(-\pi/2)Y_{2}(-\pi/2)[1/2J]$.
The time order is from right to left. It is easy to prove that

$U_{1}=X_{1}(\pi)X_{2}(\pi)Y_{1}(-\pi/2)Y_{2}(-\pi/2)$,

$U_{2}=Y_{1}(\pi/2)Y_{2}(\pi/2)$,
and

 $U_{3}=X_{1}(\pi/2)X_{2}(\pi/2)$.

 In our experiments, each NMR spectrum of spin k is obtained by a
 spin-selective readout pulse $[-\pi/2]_{y}^{k}$. The relative
 phases of signals are meaningful because all experiments are
 acquired in an identical fashion [13]. At first, we prepare the
 four pseudo- pure states described in equations(9)-(12).
 For the system in a pseudo- pure state, only one
 NMR peak appears in a spectrum if a selective readout
 pulse is applied. Fig.1 shows the experimental MNR spectra when the system
 lies in various pseudo- pure states. Figs.1a, b,
 c and d are spectra when the system lies in pseudo- pure
 states $|\uparrow\uparrow>,|\uparrow\downarrow>,|\downarrow\uparrow>$ and
$|\downarrow\downarrow>$, respectively. In each
 figure, the main spectrum represents the spectrum of $^{13}C$
  and the $^{1}H$ spectrum is a smaller inset. For each experiment,
  the initial state $|\gamma>$ is pseudo- pure
  state $|\uparrow\uparrow>$. U is selected as $U_{1},U_{2}$
  or $U_{3}$. The searching state $|\tau>$ can be any pseudo- pure
  state. $UQ$ transforms state $|\gamma>$ into
  state $|\tau>$. Therefore, the number of application of Q is
  1. The searching results on condition that
 $U=U_{3}$ and $|\tau>=|\uparrow\uparrow>$, $|\uparrow\downarrow>$,
 $|\downarrow\uparrow>$, or $|\downarrow\downarrow>$
  are shown in Fig.2. Comparing Figs.2a, b, c
  and d with Figs.1a, b, c and d respectively, we confirm that
  the system truly lies in the target state. In order
  to illustrate that the searching results for $U_{3}$ are the same
  as $U_{1}$ or $U_{2}$, Fig.3 shows the searching
  results on condition that $|\tau>=
 |\uparrow\downarrow>$, and $U=U_{1}$ (shown by Fig.3a) or $U_{2}$ (shown by Fig.3b).
 It can be found from various spectra that
 the experimental errors are not lager than 5 percent expect only two $^{1}H$ spectra.
  The errors mainly result from the imperfection of pulses,
  inhomogeneity of magnetic field and effect of decoherence.
  We find that if each $[\pi/2]_{\phi}^{k}$ ($\phi$=x or y)
  in $I_{i}$ is replaced by $[-\pi/2]_{\phi}^{k}$,
  and each $[-\pi/2]_{\phi}^{k}$ is
  replaced by $[\pi/2]_{\phi}^{k}$,
   the results  remain the same. This fact can be used
   to simplify pulse sequences and reduce experimental errors.

  In conclusion, we demonstrate the generalized quantum searching
  algorithm by replacing W-H transform by different
transformations. Because the number of repetition of operator Q is
determined by the element $|U_{\tau\gamma}|$ and these three U
transformations have the same $|U_{\tau\gamma}|$, it is not
surprised that the numbers of repetitions of Q are all 1. Compared
with temporal labelling, the spatial averaging used to prepare
initial state shortens experiment time. It only takes about 2
minutes to finish one experiment. The searching results can be
directly read out from spectra, so that the steps of recording
areas of peaks are avoided. These facts simplify the process of
experiments and make the searching results easy to observe.

 This work was partly supported by the National Nature Science Foundation
of China. We are also grateful to Professor Shouyong Pei of
Beijing Normal University for his helpful discussions on the
principle of quantum algorithm and also to Dr. Jiangfeng Du of
University of Science and Technology of China for his helpful
discussions on experiment.
\newpage
\bibliographystyle{article}

\newpage
{\begin{center}\large{Figure Captions}\end{center}
\begin{enumerate}
\item The NMR spectra of $^{13}C$(main figures)
and $^{1}H$ (smaller insets), when the two-spin system lies in
various pseudo- pure states via readout pulses selective for
$^{13}C$ (main figures) and $^{1}H$(smaller insets). The amplitude
has arbitrary units. When the system lies in a pseudo-pure state,
only one NMR peak appears in the spectrum of spin k if a readout
pulse $[-\pi/2]_{y}^{k}$ is applied. Figs.1a, b, c and d are
spectra corresponding to pseudo- pure states
$|\uparrow\uparrow>,|\uparrow\downarrow>,|\downarrow\uparrow>$and
$|\downarrow\downarrow> $, respectively.
\item Spectra of $^{13}C$ (main figures) and $^{1}H$ (smaller insets)
after completion of the generalized searching algorithm and
readout pulses selective for $^{13}C$ (main figures) and $^{1}H$
(smaller insets). U is chosen as $U_{3}$, and the target state
$|\tau>$ is
$|\uparrow\uparrow>,|\uparrow\downarrow>,|\downarrow\uparrow>$ or
$|\downarrow\downarrow>$. The corresponding searching results are
shown by Figs.2a, b, c or d. By comparing Figs.2a, b, c and d with
Figs.1a, b, c and d respectively, we confirm that the system is
truly in the target state.
\item  Spectra of $^{13}C$ (main figures) and $^{1}H$
(smaller insets) after completion of the generalized searching
algorithm and readout pulses selective for $^{13}C$ (main figures)
and $^{1}H$ (smaller insets) on condition that
$|\tau>=|\uparrow\downarrow>$, and $U=U_{1}$ shown by Fig.3a or
$U=U_{2}$ shown by Fig.3b.
\end{enumerate}
\begin{figure}{1}
\includegraphics[]{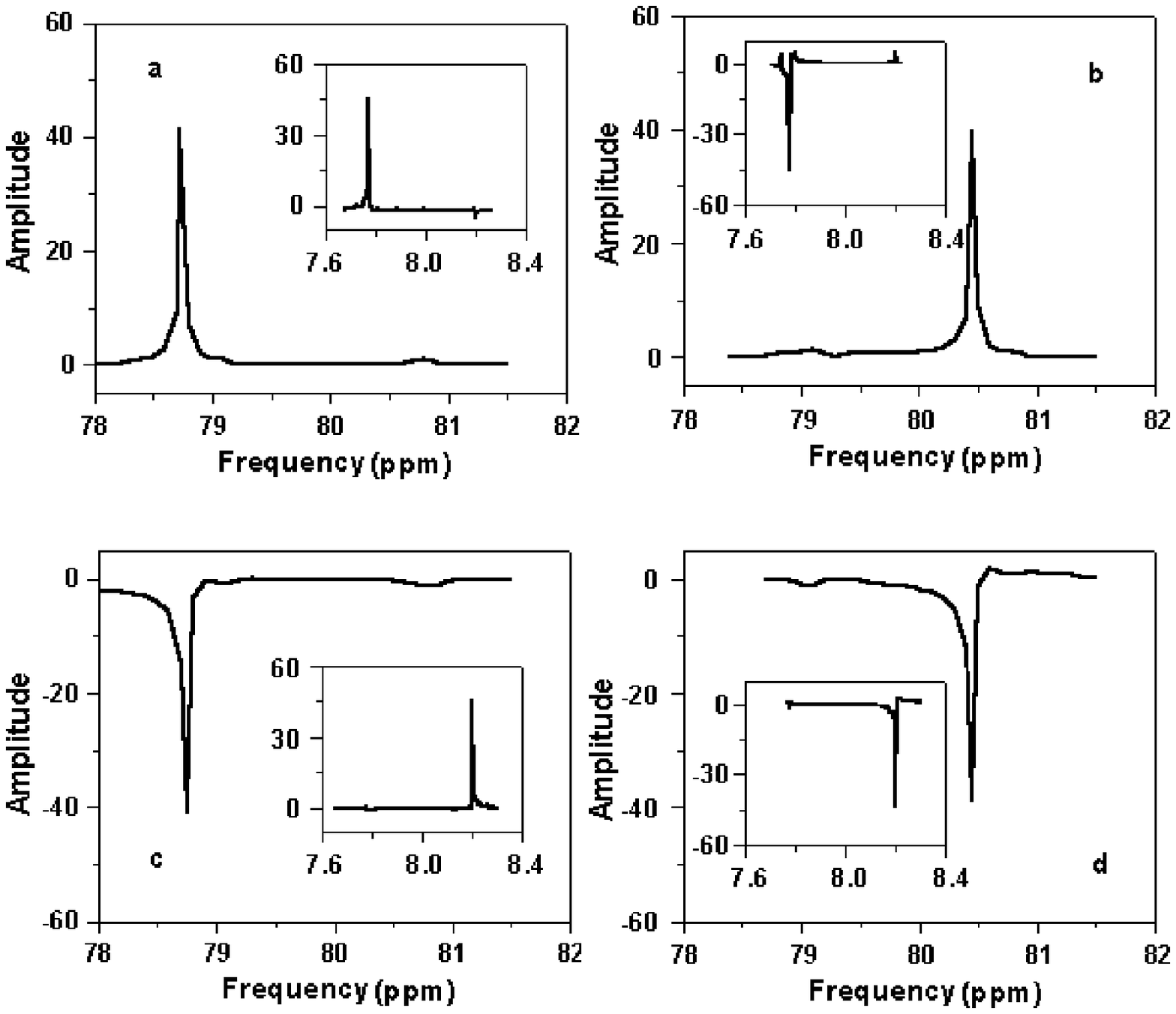}
\caption{}
\end{figure}
\begin{figure}{2}
\includegraphics[]{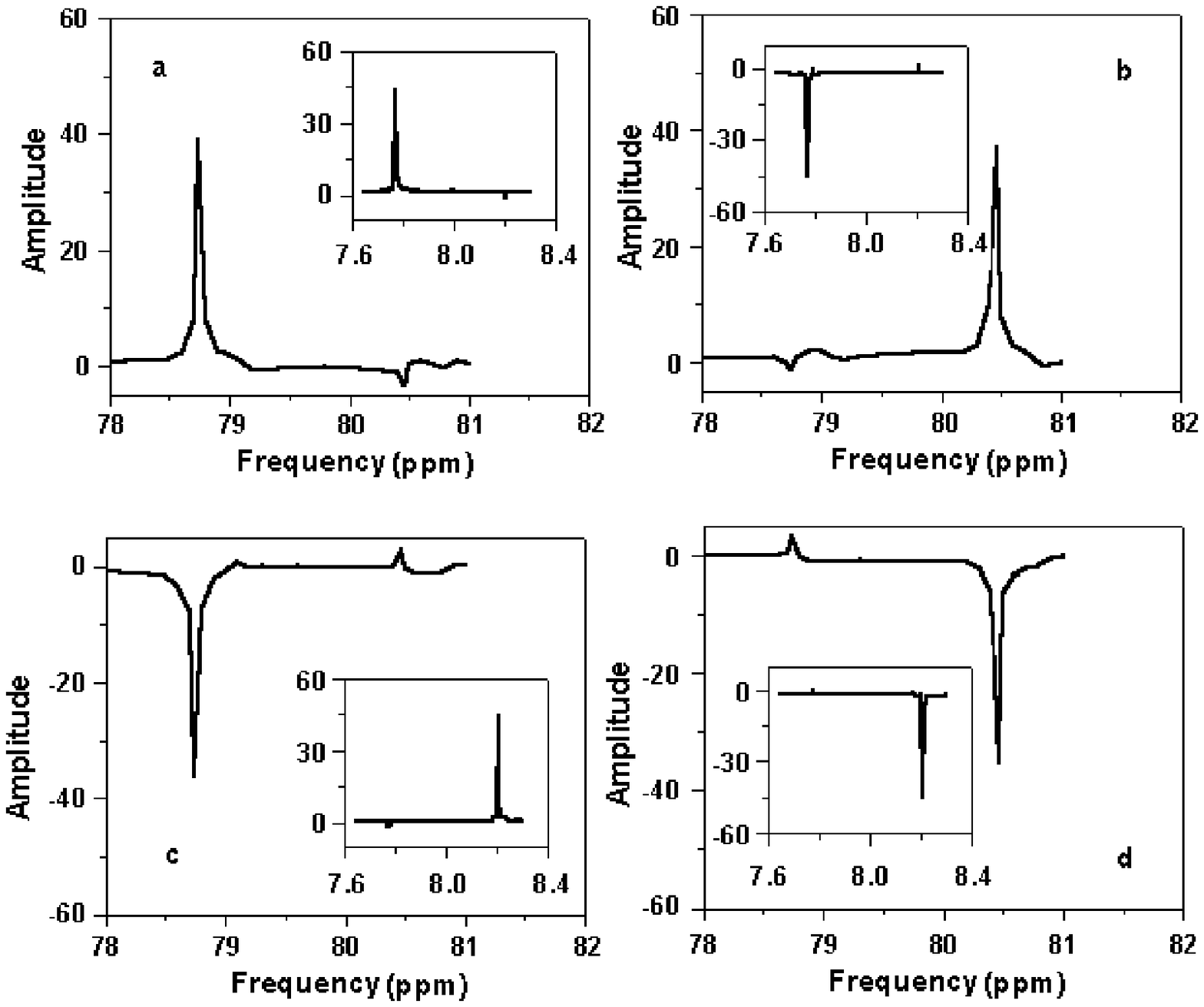}
\caption{}
\end{figure}
\begin{figure}{3}
\includegraphics[]{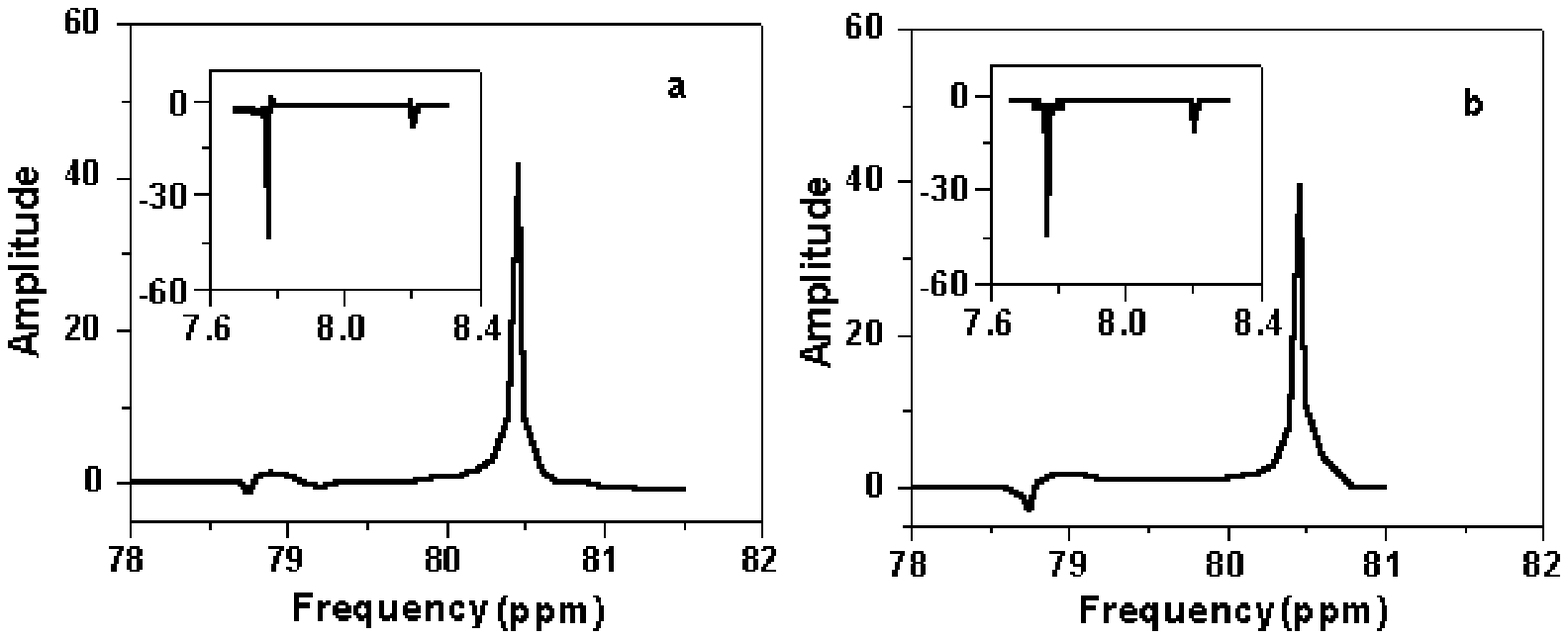}
\caption{}
\end{figure}
\end{document}